\documentstyle[epsfig,sprocl]{article}

\newcommand{\sumint}{\hbox{$\sum$}\!\!\!\!\!\!\int}
\newcommand{\Tr}{\mbox{Tr}}

\newcommand{\del}{\partial}
\newcommand{\beq}{\begin{eqnarray}}
\newcommand{\eeq}{\end{eqnarray}}
\newcommand{\be}{\begin{eqnarray*}}
\newcommand{\ee}{\end{eqnarray*}}

\newcommand{\ra}{\rightarrow}

\newcommand{\ex}[1]{\langle\,#1\,\rangle}

\def\square{\vcenter{\vbox{\hrule height.4pt
          \hbox{\vrule width.4pt height6pt
          \kern6pt\vrule width.4pt}\hrule height.4pt}}}
\def\boxx{\square}

\begin{document}

\title{RADIATIVE CORRECTIONS TO THE STEFAN-BOLTZMANN LAW}
\author{FINN RAVNDAL\footnote{Contributed talk at
        {\it Strong and Electroweak Matter '97}, Eger, Hungary, May 21-25, 1997}}
\address{Institute of Physics, University of Oslo, N-0316 Oslo, Norway}
\maketitle

\begin{abstract} Photons in blackbody radiation have non-zero interactions
due to their couplings to virtual electron-positron pairs in the vacuum. For
temperatures much less than the electron mass $m$ these effects can be described
by an effective theory incorporating the Uehling and Euler-Heisenberg interactions
as dominant terms. By a redefinition of the electromagnetic field, the Uehling 
term is shown not to contribute. The Stefan-Boltzmann energy is then 
modified by a term proportional with $T^8/m^4$ in agreement with the semi-classical 
result of Barton. The same effects give a speed of light smaller than one 
at non-zero temperature as has also recently been derived using full QED.
\end{abstract}

When one calculates the free energy of the quark-gluon plasma in QCD, one 
encounters infrared divergences in higher order perturbation theory. These
can be cured by resummation as shown by Kastening and Zhai \cite{KZ}. A more direct
derivation of the same results was subsequently given by Braaten and Nieto
using a dimensionally reduced effective theory in three dimensions \cite{BN}. At low
temperatures QCD is a strongly interacting theory and the free energy cannot
be calculated in perturbation theory. But at very low energies the relevant
hadronic degrees of freedom are pions described by a non-linear
$\sigma$-model. Even though the theory is formally non-renormalizable, radiative
corrections can be systematically calculated as long as one works at sufficiently
low energies. Using this effective field theory Gerber and Leutwyler could then calculate 
perturbative corrections to the free energy of interacting pions \cite{GL}.

Using the effective field theory approach developed by Braaten and Nieto 
for QCD, Andersen \cite{Jens} has shown that the thermodynamics of the electron-photon 
plasma of QED at high temperatures can also be derived in a more compact
way than using the alternative method of resummation \cite{Parwani}$^{\!,\,}$\cite{KZ}. 
But in contrast with QCD, one can
use QED perturbatively also at temperatures less than the electron mass. Since
only the photonic degrees are then excited, one can instead of full QED use
the effective Euler-Heisenberg theory to derive the relevant thermodynamics.
This approach goes back at least to a contribution by Tarrach \cite{Tarrach}.
Since then it has been used by Barton \cite{Barton} using a semi-classical method
to derive the first radiative correction to the energy density of blackbody 
radiation. Tarrach wanted to calculate the effects of quantum fluctuations 
on the propagation of light  at non-zero temperatures. An improved result  
by LaTorre, Pascual and Tarrach \cite{LaTorre} using full QED has since then been 
derived.
We will here consider the effective theory of low-energy photons and use it
to derive the above physical consequences. It is shown to represent a large
reduction of calculational effort. 

Integrating out the electron field in the QED partition function, we obtain the
effective Lagrangian
\be
     {\cal L}_{eff}(A) = -{1\over 4}F_{\mu\nu}^2 - i\Tr\log{[\gamma^\mu(i\del_\mu - eA_\mu) - m]}
\ee
At low energies it can be expanded in powers of momenta and gives to leading order
\be
     {\cal L}_{eff} = -{1\over 4}F_{\mu\nu}^2 + {\cal L}_U + {\cal L}_{EH} + \ldots
\ee
where the first term
\be
     {\cal L}_U = {\alpha\over 60\pi m^2}F_{\mu\nu}\boxx\, F^{\mu\nu} 
\ee
is the Uehling interaction due to the lowest-order vacuum polarization loop where $\alpha = e^2/4\pi$
is the fine structure constant and $\boxx \equiv \del_\mu\del^\mu$. The next term
\be
    {\cal L}_{EH} =  {\alpha^2\over 90m^4}\left[(F_{\mu\nu}F^{\mu\nu})^2 
                    + {7\over 4}(F_{\mu\nu}\tilde{F}^{\mu\nu})^2\right] 
\ee
is the Euler-Heisenberg interaction where 
$\tilde{F}_{\mu\nu} = {1\over 2}\epsilon_{\mu\nu\rho\sigma}F^{\rho\sigma}$ 
is the dual field strength. 

The equation of motion for the free field is $\boxx\, F_{\mu\nu} = 0$ 
and thus we see that the  Uehling term can be effectively set equal to zero as 
long as there is no matter 
present. Since the equation of motion is only satisfied by on-shell photons, one may then question 
the validity of this simplification where the interactions are used to generate loop diagrams 
with virtual photons. But since one is in general allowed to shift integration variables in the
corresponding functional integrals, we see that under the transformation
\be
     A_\mu \ra A_\mu + {\alpha\over 30\pi m^2}\boxx A_\mu 
\ee
the Uehling interaction is again removed. We are thus left with the effective Lagrangian
\be
    {\cal L}_{eff}(A) = -{1\over 4}F_{\mu\nu}^2 
                      + {\alpha^2\over 90m^4}\left[(F_{\mu\nu}F^{\mu\nu})^2 
                    + {7\over 4}(F_{\mu\nu}\tilde{F}^{\mu\nu})^2\right]
\ee
describing interacting photons at low energies in the absence of matter.

The energy density of free photons at non-zero temperature $T$ is given by the Stefan-Boltzmann law
${\cal E} = {\pi^2\over 15}T^4$. From the above effective theory we can now calculate the first         quantum correction to this classical result. For dimensional reasons it must thus vary with the 
temperature like $T^8/m^4$. Its magnitude is most directly calculated using the Matsubara
formalism where the photon field is periodic in imaginary time. In lowest order perturbation theory
the correction to the free energy density is then obtained directly from the
partition function as
\be
     \Delta {\cal F} = - {\alpha^2\over 90m^4}\ex{7F_{\mu\nu}F_{\alpha\beta}F^{\mu\alpha}F^{\mu\beta}
   - {5\over 2}F_{\mu\nu}F_{\alpha\beta}F^{\mu\nu}F^{\alpha\beta}}
\ee
where we have rewritten the term involving the dual field strength tensor $\tilde{F}_{\mu\nu}$ 
by products of $F_{\mu\nu}$ alone. In order to evaluate this expectation value we only need the
correlator
\be
     \ex{F_{\mu\nu}(K)F_{\alpha\beta}(-K)} = {1\over K^2}
     \left[K_\nu K_\alpha\delta_{\mu\beta} - K_\nu K_\beta\delta_{\mu\alpha}   
     + K_\mu K_\beta\delta_{\nu\alpha} - K_\mu K_\alpha\delta_{\nu\beta}\right]    
\ee
Using Wick's theorem we see that the result is given by the 2-loop Feynman diagram in 
Figure 1 which is really a product of two 1-loop diagrams. It is therefore almost trivial to
evaluate compared with the corresponding 3-loop diagram in full QED. One then obtains
\begin{figure}[htb]
      \begin{center}
           \epsfig{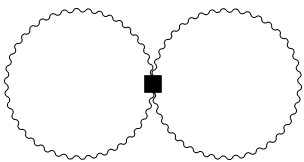}
      \end{center}
      \caption{\footnotesize Lowest order diagram for the photon
                           free energy in the effective theory.}   
      \label{double}
\end{figure}
\be
    \Delta {\cal F} = - {22\alpha^2\over 45 m^4}\sumint_P \sumint_Q \frac{(PQ)^2}{P^2Q^2}
\ee
where the basic sum-integral is
\be
    \sumint_Q {Q_\mu Q_\nu \over Q^2}  = {\pi^2 T^4\over 90}                    
    (\delta_{\mu\nu} - 4\delta_{\mu 4}\,\delta_{\nu 4})
\ee
using dimensional regularization for the space integrals and zeta-function regularization for the
Matsubara summations \cite{AB}. We thus obtain \cite{KR}
\be
    \Delta{\cal F} =  - {22\pi^4\alpha^2 \over 125\cdot 243} {T^8\over m^4} 
\ee
This result has previously been derived by Barton \cite{Barton} using a semi-classical method and
treating the interacting photon gas as a material medium. It gives directly the pressure in the
gas as $P = -{\cal F}$. The entropy is given by the derivative with respect to temperature and thus
the energy density follows from ${\cal E} = (1 - T\del/\del T){\cal F}$ as
\be
    {\cal E} = {\pi^2\over 15}T^4 + {154\pi^4\alpha^2 \over 125\cdot 243} {T^8\over m^4}  
\ee
Obviously the corrections due to the Euler-Heisenberg interaction are negligible for ordinary 
temperatures. Barton \cite{Barton} has also investigated the implications of the photon 
interactions for the Planck distribution of blackbody radiation.

It is now straightforward to calculate also the finite-temperature photon self-energy 
$\Pi_{\mu\nu}(K)$ which will follow from the Feynman diagram in Figure 2 using the above effective
Euler-Heisenberg theory. At zero temperature it has previously been considered by 
Halter \cite{Halter} who 
\begin{figure}[htb]
      \begin{center}
           \epsfig{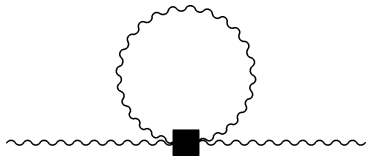}
      \end{center}
      \caption{\footnotesize Lowest order diagram for the 
                    photon self energy in the effective theory.}   
      \label{self}
\end{figure}
showed that the photon remains massless as expected. Projecting out the longitudinal and
transverse components in the formalism of Weldon \cite{Weldon}, we then obtain \cite{KR}
\be
      \Pi_L(K) = {44 \pi^2\alpha^2\over 2025}{T^4\over m^4}(K_4^2 + {\bf K}^2)
\ee
and
\be
      \Pi_T(K) = {44 \pi^2\alpha^2\over 2025}{T^4\over m^4}(-K_4^2 + {\bf K}^2)
\ee
Transforming these results back to Minkowski space, we can then obtain the permittivity $\epsilon$
and susceptibility $\mu$ of the photon vacuum at finite temperatures. In the static limit then 
follows 
\be
     \epsilon = \mu = 1 + {44 \pi^2\alpha^2\over 2025}{T^4\over m^4}
\ee
The speed of light at non-zero temperatures is therefore
\be
     c  = 1 - {44 \pi^2\alpha^2\over 2025}{T^4\over m^4}       \
\ee
It is reduced below its ordinary vacuum value due to the interaction of the photons with virtual
electron-positron pairs as first calculated by Tarrach \cite{Tarrach}. A corrected result has since
then been derived in full QED from 2-loop diagrams \cite{LaTorre} and agrees with the above result
obtained more directly in the effective theory.

\end{document}